\documentclass[]{spie}  %>>> use for US letter paper
\usepackage{amsmath,amsfonts,amssymb}
\usepackage{graphicx}
\usepackage[colorlinks=true, allcolors=blue]{hyperref}
\usepackage{tikz}
\usepackage{booktabs}
\usepackage{siunitx}
\usepackage{subcaption}
\usepackage{xspace}

\title{Design for Enabling Echelle and OMR Spectrograph Observations for Point and Extended Sources on the Same Night at VBT}

\author[a]{Nitish Singh}
\author[a]{S. Sriram}
\author[a]{Bharat Kumar Yerra}
\author[a]{Prasanna G. Deshmukh}
\affil[a]{Indian Institute of Astrophysics, II Block, Koramangala, Bengaluru 560 034, INDIA}

\authorinfo{Further author information: Residence: Bhaskara, Indian Institute of Astrophysics Guest House, Koramangala II Block, Bengaluru 560 034, India\\
Nitish Singh: E-mail: nitish.singh@iiap.res.in, nitiphy@gmail.com\\
Telephone: +91 8077759357}

\begin{document} 
\maketitle

\begin{abstract}
The 2.34m Vainu Bappu Telescope (VBT) is a reflecting telescope that operates in two modes, prime focus and cassegrain focus, and is equipped with two instruments. In prime focus mode, the telescope has the F-number of f/3.25, and the High-Resolution Echelle Spectrograph (HRES) is employed through optical fiber. On the other hand, in cassegrain focus mode, the F-number is f/13, and the OMR Spectrograph (OMRS) is mounted for low and medium-resolution spectroscopy. Currently, the VBT faces a limitation: either the OMRS or the HRES can be used due to the switch in the heavy secondary mirror. To overcome this, we present a novel method enabling the OMRS to operate from prime mode alongside the HRES. The fiber setup for OMRS is optimized with a 25-lenslet + fiber-based Integral Field Unit (IFU) capable of observing both point and extended sources. The optimized lenslet, fiber, and fore optics design is undergoing lab testing. Our approach allows seamless operation of both spectrographs on the same night, enhancing the observational capabilities of astronomical studies with VBT.
\end{abstract}

% \keywords{Prime Focus, cassegrain focus, Fiber-fed system, Integral Field Unit, Dual spectrograph operation}


\section{INTRODUCTION}
\label{sec:intro}  % \label{} allows reference to this section

Over the past few decades, newer and more efficient telescopes have been built worldwide at sites with good seeing conditions. When discussing older-generation telescopes, it becomes imperative to focus on enhancing their efficiency and integrating new technologies to keep them competitive and relevant in modern astronomy.  The Vainu Bappu Observatory (VBO) is home to 0.5 to 2-m class telescopes, including the Vainu Bappu Telescope (VBT). The VBT is a 2.34m reflecting telescope that comprises parabolic primary and hyperbolic secondary mirror systems. 
% The primary mirror has a parabolic shape, while the secondary mirror has a hyperbolic shape. 
The VBT observed its first light on November 2, 1985 (\citenum{1992BASI...20..319B}). 

Currently, the VBT is attached with two instruments, the High-Resolution Echelle spectrograph (HRES) and the OMR Spectrograph (OMRS), at prime and cassegrain focus, respectively. In prime focus mode, the telescope has an F-number of f/3.25, providing an image scale of 27 arcsec/mm. The HRES, fed through a f/3 optical fiber from prime focus, provides high spectral resolutions (R $\approx$ 27,000, 72,000, and 100,000) through different slit widths (full slit open, $60 \mu m$, $30 \mu m$) (\citenum{2005JApA...26..331R}). In cassegrain focus mode, the telescope F-number is f/13, offering an image scale of 6.7 arcsec/mm. The OMRS is directly attached to the cassegrain focus, with light being directly fed through the slit. The total effective focal length of the spectrograph is 1000 mm. Two cameras are available for OMRS: a long camera with a focal length of 420 mm and a short camera with a focal length of 150 mm. In the present OMRS setup, the slit opening for the short camera is set to 300 $\mu$m, while for the long camera, it is set to 114 $\mu$m. The maximum spectral resolution achieved using the short camera of the OMRS is approximately R$\approx$2000, and from the long camera is approximately R$\approx$5200 (\citenum{1998BASI...26..383P}). 

The VBT is dedicated to full-time spectroscopic observations through HRES and OMRS. However, the current operational setup requires switching between prime and cassegrain modes by changing the secondary mirror, which limits the usage of both instruments on the same night. The weight of the secondary mirror, approximately 750 kg, poses practical difficulties in frequent mode changes, reducing the available spectroscopic observation time with the two instruments. In this study, to increase the VBT observation efficiency, we have devised a method wherein both spectrographs can be fed through fibers from the prime focus, allowing astronomers to use both of them on the same night. The prime focus plate scale is smaller compared to the cassegrain focus, so we plan to operate both spectrographs at prime focus by increasing the field of view (FoV). We designed a wide field corrector (WFC) unit that corrects a 0.5-degree (or 72 mm) FoV at prime focus to overcome the significant aberrations in the off-axis. The telescope F number post-WFC unit is f/3.55. Since HRES is a fiber-fed system and f/3 fibers are readily available in the market, HRES can be fed through f/3 fiber from prime focus, requiring no major changes to the present setup. However, to connect OMRS from prime focus, which operates on the f/13 beam, we have designed a new lenslet+fiber-based system in ZEMAX that allows the OMRS to capture the spectra of both point and extended sources without losing any light compared to the cassegrain mode.

\section{OMRS Current Coupling Scheme at cassegrain mode}
\label{sec:current_coupl}

% The OMRS operates in cassegrain focus, utilizing two cameras: short and long with a maximum slit width for these cameras are 2 arcsec (300 $\mu$m) and 0.77 arcsec (114 $\mu$m), respectively (\citenum{1998BASI...26..383P}). The slit height remains constant at 25 mm. The plate scale in cassegrain mode is 148 $\mu$m/arcsec. The seeing conditions at the VBT site vary between 2 arcsec and 3.5arcsec, with an average seeing of 2.5 arcsec (\citenum{2009MNRAS.395..593P}). The Point Spread Function (PSF) induced by atmospheric conditions initially manifests as a speckle pattern during short exposures, evolving into a 2D Gaussian distribution with longer exposures (\citenum{1969A&A.....3..455M}). The Full Width at Half Maximum (FWHM) of the PSF aligns with the telescope’s seeing (\citenum{2023FrASS..1058213L}). To assess the flux coupling efficiency of the OMRS, I utilized a Gaussian PSF with a FWHM corresponding to the seeing conditions, which is 2.5 arcsec. I then simulated two rectangular apertures with widths of 2 arcsec and 0.77 arcsec, matching the OMRS slit dimensions for both camera cases (see Fig.~\ref{fig:omrs_coup}). By coupling the OMRS slit with this PSF, I calculated the total flux entering the spectrograph. Fig.~\ref{fig:figure2} presents the direct flux coupling of the OMRS in cassegrain mode for varying slit widths ranging from 0 to 6 arcsec and seeing conditions ranging from 2 to 3.5 arcsec.

\begin{figure}[htbp]
  \centering

  % Subfigure for Lens 1
  \begin{subfigure}[b]{0.45\textwidth}
    \centering
    \includegraphics[width=\textwidth]{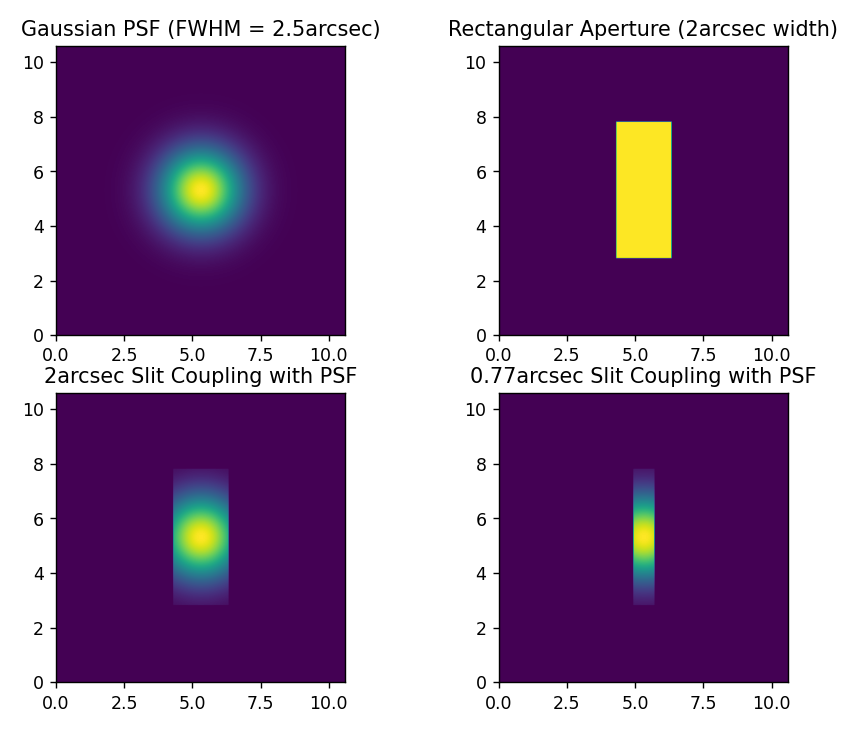}
    \caption{PSF coupling with Slit}
    \label{fig:omrs_slit}
  \end{subfigure}
  \hfill
  % Subfigure for Lens 2
  \begin{subfigure}[b]{0.52\textwidth}
    \centering
    \includegraphics[width=\textwidth]{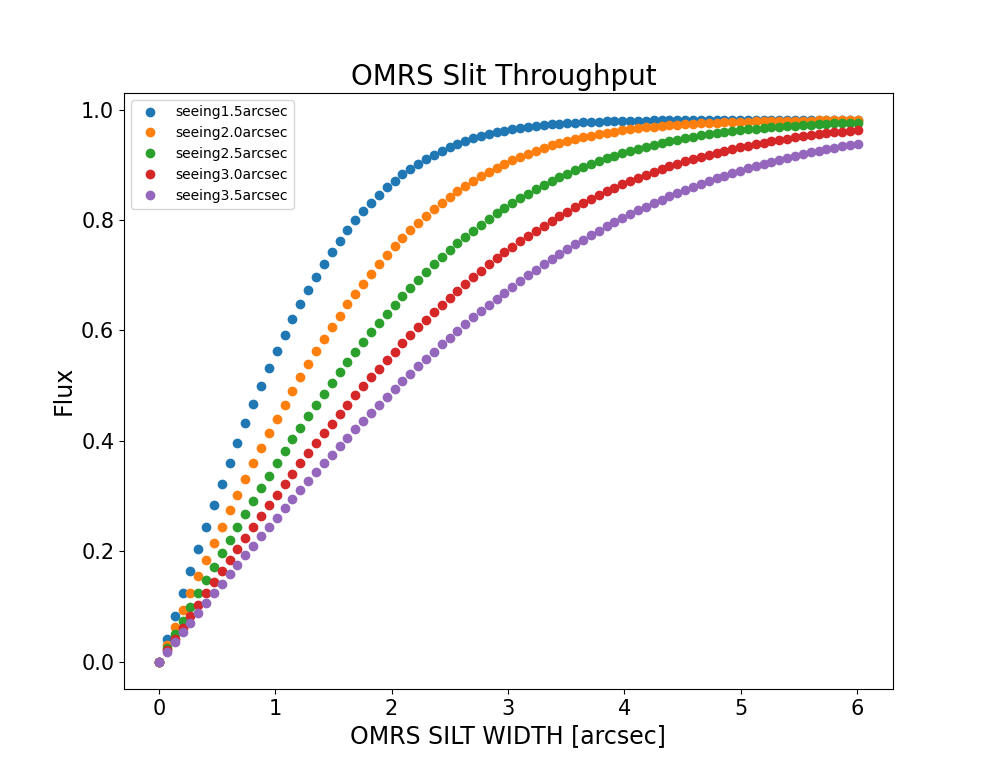}
    \caption{OMRS Slit Throughput In cassegrain Mode}

    \label{fig:omrs_coup}
  \end{subfigure}

  \caption{OMRS Current Coupling scheme at cassegrain Mode}
  \label{fig:OMRS_Cass}
\end{figure}

The OMRS has two cameras, short and long, with a maximum slit width of 2 arcsec (300  $\mu$m) and 0.77 arcsec (114 $\mu$m), respectively, and with slit height of 25 mm (\citenum{1998BASI...26..383P}). The OMRS operates in cassegrain mode at a plate scale of 148 $\mu$m/arcsec. The seeing conditions at the VBT site vary between 2 arcsec and 3.5 arcsec, with an average seeing of 2.5 arcsec (\citenum{2009MNRAS.395..593P}). The Point Spread Function (PSF) induced by atmospheric conditions initially manifests as a speckle pattern during short exposures, evolving into a 2D Gaussian distribution with longer exposures (\citenum{1969A&A.....3..455M}). The Full Width at Half Maximum (FWHM) of the PSF aligns with the telescope’s seeing (\citenum{2023FrASS..1058213L}). To assess the flux coupling efficiency of the OMRS,  Gaussian PSF with a FWHM corresponding to the seeing conditions, which is 2.5 arcsec, was used. Further, we simulated two rectangular apertures with widths of 2 arcsec and 0.77 arcsec, matching the OMRS slit dimensions for both camera cases (see Fig.~\ref{fig:omrs_slit}). By coupling the OMRS slit with this PSF, the total flux entering the spectrograph was calculated. Fig.~\ref{fig:omrs_coup} presents the direct flux coupling of the OMRS in cassegrain mode for different slit widths ranging from 0 to 6 arcsec and seeing conditions ranging from 1.5 to 3.5 arcsec.

From Fig.~\ref{fig:OMRS_Cass}: For the short camera with a slit width of 2 arcsec and a seeing of 2.5 arcsec, the coupling efficiency is 61 \%. For the long camera with a slit width of 0.77 arcsec and a seeing of 2.5 arcsec, the coupling efficiency is 27 \%.

\section{OMRS Coupling Scheme at Prime Mode}
\label{sec:new_desi}

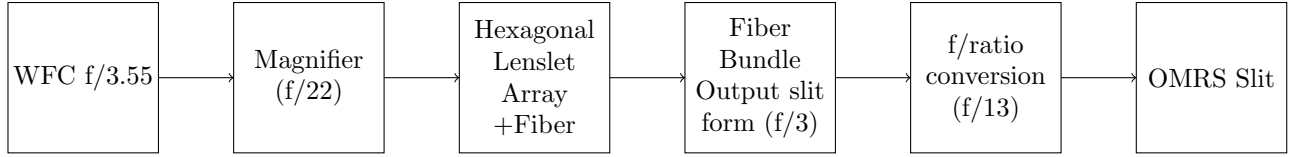
\begin{figure}[htbp]
    \centering
    \begin{tikzpicture}
        % Draw main rectangle
        \draw (-4,0) rectangle (-2,2);
        % Add labels
         \node[text width=2cm, align=center] at (-3,1){WFC f/3.55};
        % Add sub-rectangles for components
        \draw (-1,0) rectangle (1,2);
        \node[text width=2cm, align=center] at (0,1) {Magnifier (f/22)};
        \draw (2,0) rectangle (4,2);
        \node[text width=2cm, align=center] at (3,1) {Hexagonal Lenslet Array +Fiber};
        \draw (5,0) rectangle (7,2);
        \node[text width=2cm, align=center] at (6,1) {Fiber Bundle Output slit form (f/3)};

        % Add sub-rectangles for components
        \draw (8,0) rectangle (10,2);
        \node[text width=2cm, align=center] at (9,1) {f/ratio conversion (f/13)};
        \draw (11,0) rectangle (13,2);
        \node[text width=2cm, align=center] at (12,1) {OMRS Slit};

        % Add arrows
        \draw[->] (-2,1) -- (-1,1);
        \draw[->] (1,1) -- (2,1);
        \draw[->] (4,1) -- (5,1);
        \draw[->] (7,1) -- (8,1);
        \draw[->] (10,1) -- (11,1);

    \end{tikzpicture}
    \caption{ Setup of OMRS linked in Prime Focus}
    \label{fig:optical_setup}
\end{figure}

To operate the OMRS at prime focus, we designed a setup as illustrated in Fig.~\ref{fig:optical_setup}. As discussed in section~\ref{sec:intro}, following the utilization of the WFC at the prime focus, the f/ratio becomes f/3.55, and the plate scale becomes 40 $\mu m$/arcsec. The diameter of the lenslets is relatively large (507 µm, corresponding to around 13.6 arcsec), which exceeds the typical seeing conditions at the VBT site of 2.5 arcsec. Directly using these large lenslets on the focal plane would result in collecting a FoV of around 13.6 arcsec, which exceeds the resolution limit imposed by the seeing conditions.

\begin{figure}[htbp]

    \centering
    \includegraphics[width=0.9\linewidth]{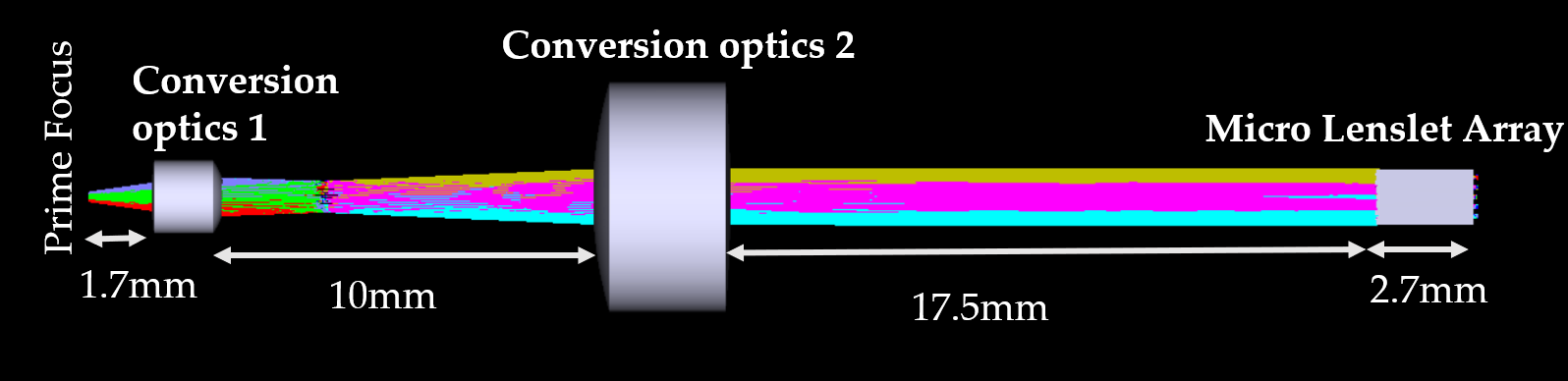}
    \caption{OMRS optical design for Prime mode with lenslet}
    \label{fig:optical_layout}

\end{figure}

 To achieve better resolution and effectively couple the seeing conditions of 2.5 arcsec, we employ a magnifier such that a 2 arcsec object becomes approximately 507 µm on the lenslet input plane, as illustrated in the Fig.~\ref{fig:optical_layout}. Subsequently, we integrate the f/3 fibers at the pupil plane of lenslet. The OMRS operates with an f/13 beam; therefore, we incorporate f/ratio conversion techniques to convert the beam from f/3 to f/13, which will be discussed further in the upcoming section \ref{sec:fiber}.

Using this setup, we can observe both point and extended sources with the OMRS. The optical design of this setup is complete (see Fig.~\ref{fig:optical_layout}) and is capable of detecting both point and extended sources.

\subsection{Selection of lenslet}
\label{sec:lenslet}

In section~\ref{sec:current_coupl}, the flux coupling decreases as the slit width is reduced (see Fig.~\ref{fig:OMRS_Cass}). From Fig.~\ref{fig:omrs_coup}, to send over 90 \% of the flux into the OMRS, we need to open the slit width to around 5 arcsec. However, doing so would not yield the required resolution according to the Rayleigh criteria if the slit width exceeds the seeing conditions. To address this issue, we employ hexagonal lenslets, with the central lenslet coupling around 2 arcsec and including the first ring of lenslets coupling around 6 arcsec, and so forth (see Fig.~\ref{fig:7lenslet_layout}). In the pupil plane of the lenslet, we will use optical fibers with sufficient separation (discussed further in Section~\ref{sec:fiber}), and each fiber will produce a spectral line on the detector. We have optimized the setup for four rings of lenslets. In our experiment, we are going to utilize f/3 optical fiber to minimize Focal Ratio Degradation with lenslet output coupled with fiber. We have to select a lenslet whose f-ratio should be between f/3 and f/4. Selecting a lenslet with a smaller diameter is crucial to mitigate Transverse Optical Spherical Aberration (TSP). The formula governing this is given by (ref: \citenum{2002PASP..114..866R})
\begin{equation}
    \text{TSP} \propto \frac{\text{Lenslet Diameter}}{\left(\text{f-ratio of lenslet} \times \text{refractive index of lenslet}^2 \right)}
\end{equation}

Regarding cross-talk, we optimize our design using various lenslets. Ultimately, we select a lenslet from Advanced Microoptic Systems GmbH company (\citenum{AuS}) with the following specifications (see Fig.~\ref{tab:microlens_spec}):

\begin{figure}[htbp]
    \begin{minipage}{0.55\textwidth}
        \centering
        \begin{tabular}{@{}ll@{}}
        \toprule
        \textbf{Parameter} & \textbf{Value} \\ 
        \midrule
        Material & Fused silica \\
        Pitch & \SI{507}{\micro\meter} \\
        Radius & \SI{0.85}{\milli\meter} \\
        Focus & \SI{1.86}{\milli\meter} (at \SI{633}{\nano\meter}) \\
        Size & \SI{10}{\milli\meter} $\times$ \SI{10}{\milli\meter} \\
        Controlling Area & \SI{9}{\milli\meter} $\times$ \SI{9}{\milli\meter} \\
        Thickness & \SI{2.7}{\milli\meter} \\
        Total number of microlenslets & $18 \times 18$ (\SI{9}{\milli\meter} $\times$ \SI{9}{\milli\meter}) \\
        Patrolling Area for our Work & $3 \times 3$ (\SI{2}{\milli\meter} $\times$ \SI{2}{\milli\meter}) \\
        F/ratio & $1.86/0.507 = 3.67$ \\
        \bottomrule
        \end{tabular}
        \caption{Specifications of the microlens array }
        \label{tab:microlens_spec}
    \end{minipage}%
    \begin{minipage}{0.45\textwidth}
    \centering
    \includegraphics[width=\textwidth]{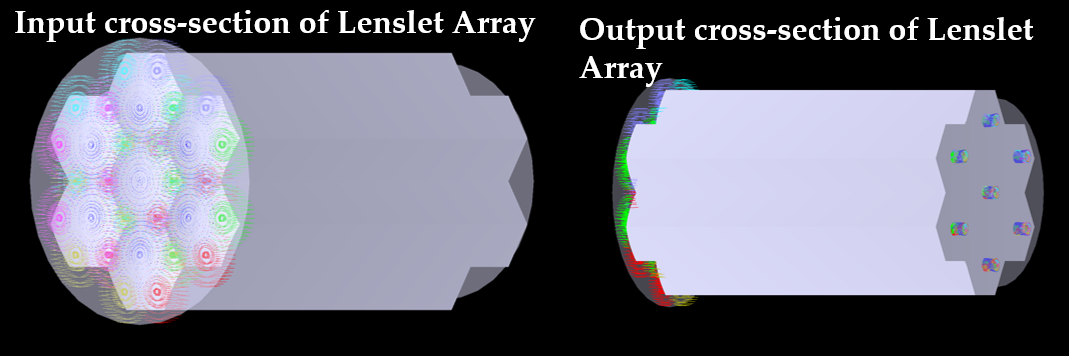}
    \caption{Optical layout of center and 1st ring of lenslets with 6 arcsec FoV}
    \label{fig:7lenslet_layout}
    \end{minipage}
\end{figure}

So, we aim to couple 2 arcsec with the center lenslet or 4 arcsec with one ring of lenslets. However, the lenslet pitch is significantly higher compared to the plate scale at the prime focus, which is $40 \mu m/arcsec$ after introducing WFC. With 2 arcsec (equivalent to $80 \mu m$), and our lenslet diameter being 507 µm, we utilize conversion optics or magnifier optics to magnify the upcoming beam such that 2 arcsec approximately equals 507 µm. We use the following formula for this calculation (ref: \citenum{2002PASP..114..866R}):
\begin{equation}
    \begin{aligned}
        \text{Conversion factor} &= \frac{\text{Diameter of lenslet ($\mu m$)}}{\text{FoV to be coupled by lenslet (µm)}} \\
        &= \frac{507}{2 \times 40} = 6.33
    \end{aligned}
\end{equation}

\begin{equation}
    \begin{aligned}
        \text{Magnification in F-ratio after Conversion optics} &= \text{Conversion factor} \times \text{F-ratio after WFC} \\
        &= 6.33 \times 3.55 = 22.4
    \end{aligned}
\end{equation}

We chose front-end converter lenses from Edmund Optics (\citenum{EdmundOptics}) to convert the f/3.55 beam to an f/22 beam. For magnifying the beam, we use two achromatic doublet lenses, Conversion Optics 1 and Conversion Optics 2, as shown in Fig.~\ref{fig:optical_layout}.

Conversion Optics 1 (L1): 2 mm Diameter x 3.0 mm Focal Length, MgF2 Coated, Achromatic Doublet Lens 

Conversion Optics 2 (L2): 6.25 mm Diameter x 17.5 mm Focal Length, VIS-NIR, Inked, Achromatic Lens 

The conversion factor is calculated as the ratio of the focal length of L2 to the focal length of L1, resulting in a value of 8.75. We observe that the conversion factor obtained is large, as when we feed this lens in ZEMAX, we obtain an f/ratio of f/22, which is our target.

\subsection{Crosstalk Analysis for Center Lenslet Field} 
\label{sec:crosstalk}

\begin{figure}[htbp]
  \centering

  % Subfigure for Lens 1
  \begin{subfigure}[b]{0.35\textwidth}
    \centering
    \includegraphics[width=\textwidth]{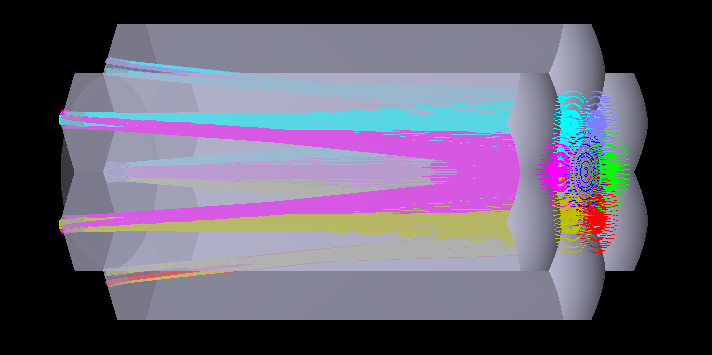}
    \caption{Optical layout of center FoV (2 arcsec distribution) with center and 1st ring of lenslets}
    \label{fig:7lenslet_centerFoV_layout}
  \end{subfigure}
  \hfill
  % Subfigure for Lens 2
   \centering
  \begin{subfigure}[b]{0.30\textwidth}
    \centering
    \includegraphics[width=\textwidth]{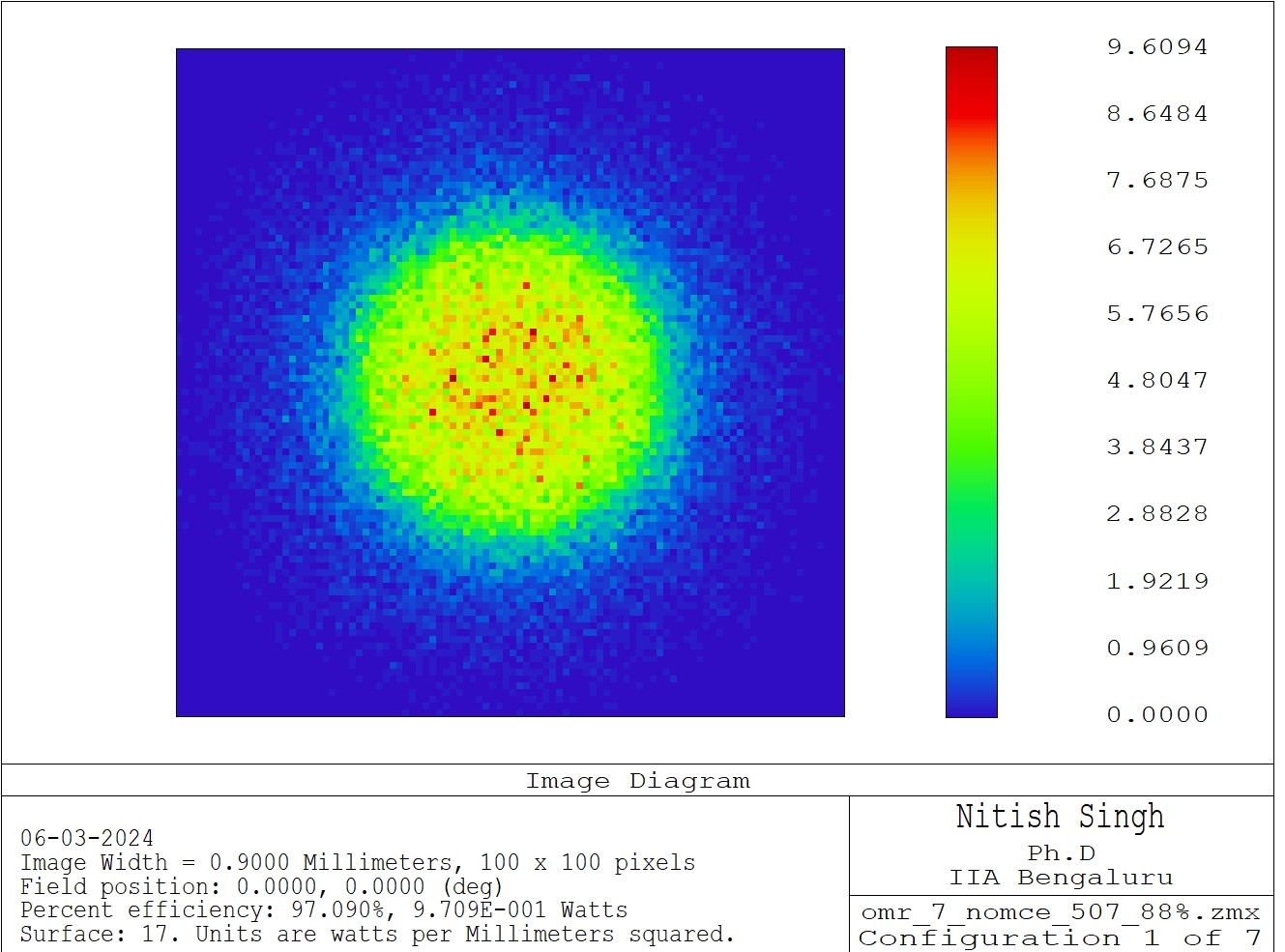}
    \caption{2 arcsec FoV total flux before lenslet plane}
    \label{fig:flux_before_lenslet}
  \end{subfigure}
  \hfill
  \begin{subfigure}[b]{0.30\textwidth}
    \centering
    \includegraphics[width=\textwidth]{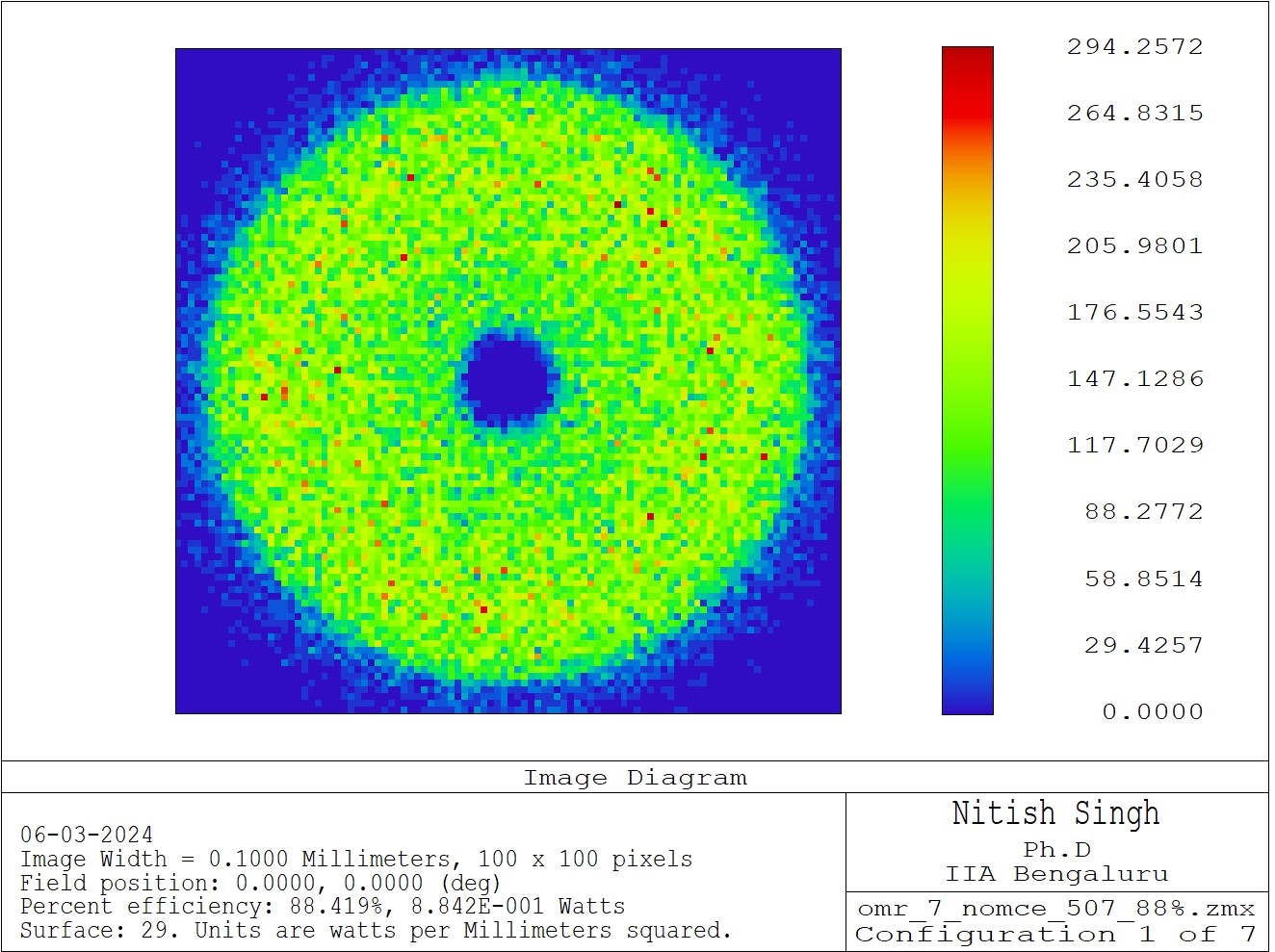}
    \caption{2 arcsec FoV total flux couple by the center lenslet}

    \label{fig:flux_after_lenslet}
  \end{subfigure}
  \caption{Cross Talk Calculation for center lenslet}
  \label{fig:cross_talk}
\end{figure}

 To calculate cross-talk, we analyze the magnified incoming beam that couples with the lenslet array. Specifically, for the center lenslet targeting a 2 arcsec field (as illustrated in Fig.~\ref{fig:7lenslet_centerFoV_layout}), we evaluate the incident flux before the lenslet and quantify the resulting flux at the pupil of the center lenslet. In the Fig.~\ref{fig:flux_before_lenslet}, a 2 arcsec field is provided, and it is determined that around 97 \% of the flux is incoming. At the pupil of the center lenslet, a maximum of 88 \% of the flux is observed in Fig.~\ref{fig:flux_after_lenslet}. Given the diameter is 90 µm, we can collect a total of 88 \% of the flux. However, we opt for a 100 µm diameter fiber to account for alignment errors.

This implies that approximately 9$\%$  of the flux either goes nearby the lenslet or deviates. Subsequently, we investigated how much flux is diverted to other lenslets within the 2 arcsec FoV. It was observed that around 1.4 $\%$ of the flux is directed to each nearby lenslet of the center lenslet. 

Therefore, the total cross-talk is calculated as (6 $\times$ 1.4 = 8.4 \%) of the flux:

\hspace{1mm}
\begin{equation}
 \text{Total cross-talk} = 2 \text{arcsec} \times 0.084 = 0.168 \text{arcsec}
\end{equation}

\subsection{Optimizing Fiber Density for Spectrograph Integration}

\label{sec:fiber}

In the pupil plane of the lenslet, we plan to use a 100 $\mu$m fiber, which will output an f/3 beam. However, the OMRS operates with an f/13 beam. To bridge this gap, we will integrate two specific Edmund Optics lenses (\citenum{EdmundOptics}) to convert the f/3 beam to an f/13 beam. Below are the details of the lenses:

Conversion Optics 3 (L3): 25 mm Dia. x 35 mm FL, VIS-NIR, Inked, Achromatic Lens

Conversion Optics 4 (L4): 25 mm Dia. x 150 mm FL, VIS-NIR, Inked, Achromatic Lens

To convert the f/3 beam to an f/13 beam, we employ the following formula from (ref: \citenum{2002PASP..114..866R}):
\begin{equation}
    \text{Beam Diameter after L3 lens (BD3)} = \frac{\text{Focal length of L3}}{\text{f/ratio}} = \frac{35}{3} = 11.6 \text{mm}
\end{equation}

\begin{equation}
    \text{f/ratio after L4} = \frac{\text{Focal length of L4}}{\text{Beam diameter of L3}} = \frac{150}{11.6} = 12.93
\end{equation}

Now, the size of the object falling on the OMRS slit will be calculated using the formula provided in (ref: \citenum{2002PASP..114..866R}).

\begin{equation}
    \text{Size of the object} = \frac{\text{Core diameter of fiber} \times \text{f/13}}{\text{f/ratio of the fiber bundle output}}
\end{equation}

\begin{equation}
    \text{Size of the object} = 100\mu \text{m} \times \frac{13}{\text{3}} = 433\mu \text{m}
\end{equation}

Therefore, the object's size that will pass through the OMRS slit from the fiber bundle is approximately 433 $\mu$m. 

\begin{equation}
    \text{New FWHM which will go inside the Slit} = 433\mu \text{m} = \frac{433 \times 3.55}{40 \times 13} arcsec = 2.9 arcsec  
 \end{equation}

So now, the OMRS slit height is 25 mm. We need at least a 5 $\sigma$ gap between two fibers to prevent overlap on the CCD. 

\begin{equation}
    5 \sigma = \frac{5 \times 433}{2.35} = 921\mu \text{m}
\end{equation}

Between two fibers, we need around a 921 $\mu$m gap, and the total height of the slit is 25000 $\mu$m.

\begin{equation}
    \text{Total number of fibers that can be accommodated} = \frac{25000}{921} = 27
\end{equation}

So, approximately, we can feed 27 fibers inside the OMRS slit while maintaining sufficient separation between spectra on the detector. However, we aim to accommodate more than 30 fibers by reducing the fiber diameter. Considering the total number of fibers fed into the OMRS slit, we plan to utilize 3 to 4 rings of hexagonal lenslets. This arrangement will extend the total FoV covered by this system to approximately $12 \times 10$ arcsec$^2$.

\section{OMRS throughput at cassegrain and Prime Mode }
\label{sec:result}

For calculations, we considered the reflectivity of 90 \% for the secondary mirror and a transmittance of 98 \% for all lenses and lenslets, along with a fiber coupling and transmittance efficiency of 96 \%. We computed the slit throughput for different fiber sizes from Fig.~\ref{fig:omrs_coup}. In Table~\ref{tab:thro_cass} and ~\ref{tab:thro_prime}, we calculate the total throughput within the spectrograph, considering the average seeing condition of 2.5 arcsec at VBT.

\begin{table}[htbp]
    \centering
    \caption{Total Throughput going inside the OMRS slit at cassegrain Mode}

    \begin{tabular}{|r|p{5cm}|p{3cm}|p{3cm}|p{3cm}|p{3cm}|r|}
        \hline
        \textbf{S/No.} & \textbf{Parameter} & \textbf{Short Camera} & \textbf{Long Camera}   \\
         \hline
        1 & Secondary mirror Reflectivity  & 0.90 &  0.90  \\
        \hline

        2 & Slit Throughput  & 0.64  & 0.27    \\
        \hline       
        3 & Total Throughput in OMRS   &  0.57 & 0.24    \\
        \hline

    \end{tabular}
    \label{tab:thro_cass}
\end{table}

\begin{table}[htbp]
    \centering
    \caption{Total Throughput going inside the OMRS slit at prime mode}

    \begin{tabular}{|r|p{5cm}|p{3cm}|p{3cm}|p{3cm}|p{3cm}|r|}
        \hline
        \textbf{S/No.} & \textbf{Parameter} & \textbf{Short Camera} & \textbf{Long Camera}   \\
         \hline
        1 & Corrector efficiency  & \((0.98)^3\) &  \((0.98)^3\)  \\
        \hline

        2 & Magnifier efficiency  & \((0.98)^2\)  & \((0.98)^2\)    \\
        \hline       
        3 & Lenslet transmission efficiency   &  0.98 & 0.98    \\
        \hline
        4 & Lenslet and Fiber Coupling  & 0.96 & 0.96     \\
        \hline
       
        5 & Fiber Transmission effi.  &  0.96 & 0.96   \\
        \hline
        6 & f/ratio conversion effi.  & \((0.98)^2\) & \((0.98)^2\)    \\
        \hline
        7 & Slit Throughput  & 0.72 & 0.36     \\
        \hline
       
        8 & Total Throughput in OMRS  &  0.56 & 0.25   \\
        \hline

    \end{tabular}
    \label{tab:thro_prime}
    
\end{table}

Table~\ref{tab:thro_cass} presents the total throughput within the OMRS using the current cassegrain configuration. We observe a throughput of 64 \% for the short camera and 27 \% for the long camera. In contrast, Table~\ref{tab:thro_prime} presents the total throughput for the OMRS new design configuration optimized for operation at the prime focus. In this configuration, the slit throughput depends on the diameter of the fiber. The diameter of the fiber is determined by the size of the image formed by the lenslet on the pupil plane (see Fig.~\ref{fig:cross_talk}). The coupling efficiency of the lenslet and fiber indicates how much light the lenslet can couple after magnifying the beam and focusing it on the pupil plane, and how much light is subsequently coupled by the fiber. With our new design, we achieve a slit throughput of 72 \% for the short camera and 36 \% for the long camera. The total throughput in OMRS is comparable to the current setup when considering the transmission efficiency of all the optics used in the new design. For the current OMRS setup in cassegrain mode, the total throughput within the spectrograph is 57 \% for the short camera and 24 \% for the long camera (see Table~\ref{tab:thro_cass}).  In comparison, for the new design, the total throughput within the spectrograph is around 56 \% for the short camera and 25 \% for the long camera (see Table~\ref{tab:thro_prime}).

\section{Conclusion}

We have designed a fiber-fed system to connect the OMRS through the prime focus of the VBT alongside HRES. For HRES, we are keeping its current fiber configuration as the post-WFC beam is f/3.55. However, for the OMRS, which is currently in cassegrain mode, we are adapting it for use in prime focus. To achieve this, we have compared the performance of the OMRS in its current cassegrain setup with our new prime focus setup. In Table~\ref{tab:thro_cass} and \ref{tab:thro_prime}, we present all the calculations comparing the OMRS's current setup in cassegrain mode, which has an f/13 beam, to our new design for prime mode, which has an f/3.55 beam after the introduction of the WFC. Our new design maintains the same throughput performance as the current OMRS setup. Additionally, our OMRS fiber configuration enables observers to acquire low and medium-resolution spectra of point sources and extended sources with a FoV of up to approximately $12 \times 10$ arcsec$^2$ in size. Furthermore, the newly employed WFC unit allows the setup of the HRES Fiber on-axis and the position of the OMRS fiber off-axis, allowing both spectrographs to be utilized on the same night. This setup will enable astronomers to observe the low, medium, and high-resolution spectra of astronomical objects on the same night. 
This setup for VBT will provide multi object observational capability.
% This setup will work in prime mode alongside the HRES, providing an efficient and versatile observational capability.

% References
\bibliography{report} % bibliography data in report.bib
\bibliographystyle{spiebib} % makes bibtex use spiebib.bst

\end{document}